# Multiqubit Spin

Alexander R. Kessel and Vladimir L. Ermakov

*Physicotechnical Institute*
*Kazan Science Center of the Russian Academy of Sciences*
*Kazan 420029 Russia*
*e-mail*: ermakov@sci.kcn.ru

It is proposed that the state space of a quantum object with a complicated discrete spectrum can be used as a basis for multiqubit recording and processing of information in a quantum computer. As an example, nuclear spin 3/2 is considered. The possibilities of writing and reading two quantum bits of information, preparation of the initial state, implementation of the "rotation" and "controlled negation" operations, which are sufficient for constructing any algorithms, are demonstrated.

We shall consider a nucleus with spin I = 3/2 and with an electric quadrupole moment in an axial symmetric crystalline electric field and a constant magnetic field parallel to the symmetry axis. We assume that the Zeeman energy is greater than the quadrupole energy, so that the nucleus possesses four nonequidistant spin energy levels $E_m$ with eigenfunctions $\chi_m$ ( m = ±3/2, ±1/2), which are eigenfunctions of the z component $\mathbf{I}_z$ of the nuclear spin $\mathbf{I}$.

Let a radio frequency (RF) field $2H_a \cos\Omega_a t$, polarized along the y axis and resonant for a certain pair of energy levels $\hbar\Omega_a = E_m - E_n$, act on the spin. The interaction operator $H_a$ with the field in the interaction representation contains a time-independent term $H_{a,\,eff} = \hbar\gamma H_a[<m|I_Y|n>\mathbf{P}_{mn} + <n|I_Y|m>\mathbf{P}_{nm}]$ and rapidly oscillating (at frequencies which are sums of $\Omega_a$ and $(E_k - E_l)/\hbar$) terms. The role of these last terms in the evolution of the spin states is negligible. To simplify the notation here and below we use the projection operators $\mathbf{P}_{mn}$ representation, where $\mathbf{P}_{mn}$ is a 4x4 matrix with all matrix elements $p_{kl}$ equal to zero except $p_{mn} = 1$. Projection operators are very convenient because of their very simple properties: $\mathbf{P}_{kl}\mathbf{P}_{mn} = \delta_{lm}\mathbf{P}_{kn}$ and $\mathbf{P}_{mn}\chi_k = \delta_{nk}\chi_m$. In addition, 1,2,3, and 4 will stand for the indices –3/2, -1/2, +1/2, +3/2, respectively.

The interaction $H_{a,eff}$ leads to spin states evolution managed by propagator

$$\mathbf{U}(\varphi_a) = [\mathbf{P}_{mm} + \mathbf{P}_{nn}] \cos(\varphi_a/2) + [\mathbf{P}_{kk} + \mathbf{P}_{ll}] + [\mathbf{P}_{nm} - \mathbf{P}_{mn}] \sin(\varphi_a/2), \qquad (1)$$

where $\varphi_a = 2(\gamma H_a t)|<m|I_Y|n>|$, $E_m > E_n$ and the indices $k,l \neq m,n$. If the RF field $2H_f \cos\Omega_{mn} t$ is polarized along the x axis, the evolution operator is

$$U_f(\varphi_f) = [\mathbf{P}_{mm} + \mathbf{P}_{nn}] \cos(\varphi_f/2) + [\mathbf{P}_{kk} + \mathbf{P}_{ll}] + i[\mathbf{P}_{nm} + \mathbf{P}_{mn}] \sin(\varphi_f/2), \quad (1a)$$

where $\varphi_f = 2(\gamma H_f t)|<m|I_X|n>|$.

In the currently accepted NMR model of a quantum computer two *real* exchange coupled spins R=1/2 and S=1/2 are considered as a basis for constructing quantum logic elements [1-4]. In the formalism of quantum mechanics the states of such a system and the operations on them are written in *abstract* four dimensional space, which is a direct product $\Gamma_R \otimes \Gamma_S$ of the two-dimensional spaces of eigenstates of the *real* spins **R** and **S**. In our case, to clarify the information aspect of the proposed logical operations it is convenient to perform the inverse procedure: to represent the four-dimensional space $\Gamma_I$, corresponding to *real* spin 3/2, as a direct product $\Gamma_R \otimes \Gamma_S$ of two *abstract* two-dimensional states spaces of *virtual* spins **R** and **S**. Then any operator **P** in the four-dimensional basis can be expressed as a linear combination of the direct products **R** $\otimes$ **S** of operators given in the subspaces $\Gamma_R$ and $\Gamma_S$. The following isomorphic correspondence exists between the basis $|m>$ of the space $\Gamma_I$ and the basis $|m_1>\otimes|m_2>$ of the direct product $\Gamma_R \otimes \Gamma_S$:

$|\chi_1> \propto |-1/2>|-1/2> \equiv |11>$,    $|\chi_3> \propto |+1/2>|-1/2> \equiv |01>$,

$|\chi_2> \propto |-1/2>|+1/2> \equiv |10>$,    $|\chi_4> \propto |+1/2>|+1/2> \equiv |00>$,

where $|10>$ and so on are the notations adopted in information theory for states of two quantum bits (qubits). The energies corresponding to these states satisfy conditions: $E_1 > E_2 > E_3 > E_4$.

The initial state for known quantum algorithms suggested for an abstract quantum computer is the state $|00> \propto |\chi_4>$. From the standpoint of the quantum calculations the subsequent spin density matrix

$$\rho_{init} = \text{const}\,[\mathbf{1} + \text{const}\,\mathbf{P}_{44}], \quad (2)$$

is the equivalent of the state |00>. Here **1** is the unit matrix in the space $\Gamma_I$. It does not change under unitary computational transformations and does not contribute to the observed signal when the result is read out.

In a sample of macroscopic size a collection of such nuclei forms an ensemble, whose spin levels in equilibrium are populated according to the Boltzmann distribution

$$\boldsymbol{\rho}_{eq} = Z \exp(-\beta \boldsymbol{H}), \qquad Z^{-1} = \mathrm{Sp}[\exp(-\beta \boldsymbol{H})], \quad \beta = 1/\kappa T, \tag{3}$$

and at room temperature the relative populations difference of the spin levels is of the order of $10^{-5}$ - $10^{-6}$ or less. Therefore, to obtain the state $\boldsymbol{\rho}_{init} = \mathrm{const}\, \boldsymbol{P}_{44}$ directly by cooling one requires very low temperatures, which, besides substantial technological difficulties, will slow down the speed of the entire computational cycle.

To obtain the desired initial state for calculation we propose the procedure whose idea goes back to paper [3]. Let the required calculation consist of performing the transformation $\boldsymbol{U}_{comp}$ of the state $\boldsymbol{\rho}_{init}$, while the spin system is in a state given by the density matrix (3), which in the high-temperature approximation is

$$\boldsymbol{\rho}_{eq} = Z\left[\mathbf{1} + \sum \lambda_m \boldsymbol{P}_{mm}\right], \qquad \mathbf{1} = \sum \boldsymbol{P}_{mm} \tag{4}$$

where $\lambda_m = E_m / kT$ and $m = 1,2,3,4$. We shall show that the sum of the results of three specially designed transformations of the state $\boldsymbol{\rho}_{eq}$ is equivalent to a transformation of the state $\boldsymbol{\rho}_{init}$. Indeed, assume that we are required to perform a calculation which is a unitary transformation $\boldsymbol{U}_{comp}$ of the spin **I** states. We define the unitary transformations

$$\boldsymbol{U}_1 = \boldsymbol{U}_a(\pi)\boldsymbol{U}_b(\pi) = \boldsymbol{P}_{13} + \boldsymbol{P}_{21} + \boldsymbol{P}_{32} + \boldsymbol{P}_{44},$$

$$\boldsymbol{U}_2 = \boldsymbol{U}_b(\pi)\boldsymbol{U}_a(\pi) = -\boldsymbol{P}_{12} - \boldsymbol{P}_{23} + \boldsymbol{P}_{31} + \boldsymbol{P}_{44},$$

where $\boldsymbol{U}_a(\pi)$ is the propagator (1) for a pulsed RF field at frequency $\Omega_a = (E_1 - E_2)/\hbar$, which performs a rotation by the angle $\varphi = \pi$, and $\boldsymbol{U}_b(\pi)$ is the propagator for a resonant RF pulse at frequency $\Omega_b = (E_2 - E_3)/\hbar$. It can be verified that the average of the three transformations $\boldsymbol{U}_{comp}$, $\boldsymbol{U}_{comp}\boldsymbol{U}_1$ and $\boldsymbol{U}_{comp}\boldsymbol{U}_2$,

$$(1/3)[\mathbf{U}_{comp}\boldsymbol{\rho}_{eq} \mathbf{U}^{\dagger}_{comp} + \mathbf{U}_{comp}\mathbf{U}_1 \boldsymbol{\rho}_{eq} \mathbf{U}^{\dagger}_1\mathbf{U}^{\dagger}_{comp} + \mathbf{U}_{comp}\mathbf{U}_2 \boldsymbol{\rho}_{eq} \mathbf{U}^{\dagger}_2 \mathbf{U}^{\dagger}_{comp}] =$$

$$=(1/3)\mathbf{U}_{comp}(\boldsymbol{\rho}_{eq} + \mathbf{U}_1 \boldsymbol{\rho}_{eq} \mathbf{U}^{\dagger}_1 + \mathbf{U}_2 \boldsymbol{\rho}_{eq} \mathbf{U}^{\dagger}_2) \mathbf{U}^{\dagger}_{comp} = \mathbf{U}_{comp}\boldsymbol{\rho}_{init} \mathbf{U}^{\dagger}_{comp}$$

is equivalent to the transformation $\mathbf{U}_{comp}$ of the density matrix

$$\boldsymbol{\rho}_{init} = Z [\alpha \mathbf{1} + \beta \mathbf{P}_{44}], \tag{5}$$

where $\alpha = 1+(1/3)[\lambda_1 +\lambda_2 + \lambda_3]$, and $\beta = \lambda_4 - (1/3)[\lambda_1 +\lambda_2 + \lambda_3]$.

Let us find the propagator corresponding to the rotation by a certain angle in the space $\Gamma_S$ under the condition that the space $\Gamma_R$ is invariant. This can be done by acting on the spin $\mathbf{I}$ with a two-frequency RF pulse containing the resonance frequencies $\Omega_a = (E_1 - E_2)/\hbar$ and $\Omega_c = (E_3 - E_4)/\hbar$. The propagator of such a transformation in the space $\Gamma_I$ will be the product of propagators of the form (1) which perform rotation by the same angle $\varphi_a = \varphi_c = \varphi$ at each transition

$$\mathbf{U}_{a+c}(\varphi, \varphi) = \mathbf{1} \cos(\varphi/2) + [\mathbf{P}_{21} - \mathbf{P}_{12} + \mathbf{P}_{43} - \mathbf{P}_{34}] \sin(\varphi/2).$$

It can be expressed as follows in terms of the operators of the spaces $\Gamma_R$ and $\Gamma_S$:

$$\mathbf{U}_{a+c}(\varphi, \varphi) = (\mathbf{R}_{11} + \mathbf{R}_{22})\otimes[(\mathbf{S}_{11} + \mathbf{S}_{22}) \cos(\varphi/2) + (\mathbf{S}_{21} - \mathbf{S}_{12})\sin(\varphi/2)] =$$

$$= \exp\{i (\varphi/2) \mathbf{1}_R \otimes \mathbf{S}_Y\}, \tag{6}$$

which proves the assertion made above (for virtual spins the indices 1 and 2 are used instead of $-1/2$ and $+1/2$, respectively; unit matrices are defined as $\mathbf{1}_R = \Sigma \mathbf{R}_{mm}$ and $\mathbf{1}_S = \Sigma \mathbf{S}_{mm}$, with m = 1,2). Similarly, the propagator for a two-frequency RF pulse with carrier (filling) frequencies $\Omega_d = (E_1 - E_3)/\hbar$ and $\Omega_e = (E_2 - E_4)/\hbar$ and angles $\varphi_d = \varphi_e = \varphi$ will be equal to

$$\mathbf{U}_{d+e}(\varphi,\varphi) = [\mathbf{P}_{22} + \mathbf{P}_{44}] \cos(\varphi/2) + [\mathbf{P}_{42} - \mathbf{P}_{24}] \sin(\varphi/2) +$$

$$+ [\mathbf{P}_{33} + \mathbf{P}_{11}] \cos(\varphi/2) + [\mathbf{P}_{31} - \mathbf{P}_{13}] \sin(\varphi/2).$$

It can be expressed in terms of spaces $\Gamma_R$ and $\Gamma_S$ spin operators as

$$\mathbf{U}_{d+e}(\varphi,\varphi) = [(\mathbf{R}_{11} + \mathbf{R}_{22}) \cos(\varphi/2) + (\mathbf{R}_{21} - \mathbf{R}_{12})\sin(\varphi/2)] \otimes (\mathbf{S}_{11} + \mathbf{S}_{22}) =$$

$$= \exp\{ i (\varphi/2) \mathbf{R}_x \otimes \mathbf{1}_S\}. \qquad (7)$$

which is the rotation by the angle $\varphi$ in the space $\Gamma_R$ with $\Gamma_S$ remaining unchanged. Generally speaking, the transitions at the frequencies $\Omega_d$ and $\Omega_e$ are forbidden in the initially adopted configuration of the magnetic and crystalline electric fields. Here we shall assume that gradient electric field asymmetry parameter is not equal zero and a small entanglement of the wave functions takes place. This allows «forbidden» transitions, and a larger amplitude of the RF field provides the required rotation angle. All results of the present letter can be obtained also with an arbitrary configuration of static magnetic and electric fields, but this would unjustifiably complicate the paper.

Next, the transformation $\mathbf{U}_f(\varphi_f)$ at $\varphi_f = \pi$, defined as

$$\mathbf{U}_f(\pi) = [\mathbf{P}_{33} + \mathbf{P}_{44}] + i [\mathbf{P}_{21} + \mathbf{P}_{12}],$$

performs the two-bit operation "controlled negation" CNOT – it performs the operation NOT on spin S if the spin R is in state |1> and leaves the spin S unchanged if the spin R is in state |0>. Indeed, it is easy to check that

$$\mathbf{U}_f(\pi) |\chi_1\rangle \equiv \mathbf{U}_f(\pi) |11\rangle = |10\rangle, \qquad \mathbf{U}_f(\pi) |\chi_2\rangle \equiv \mathbf{U}_f(\pi) |10\rangle = |11\rangle,$$

$$\mathbf{U}_f(\pi) |\chi_3\rangle \equiv \mathbf{U}_f(\pi) |01\rangle = |01\rangle, \qquad \mathbf{U}_f(\pi) |\chi_4\rangle \equiv \mathbf{U}_f(\pi) |00\rangle = |00\rangle.$$

Hence one can see that the evolution operator $\mathbf{U}_f(\pi)$ can be represented in the basis $\Gamma_R \otimes \Gamma_S$ as

$$\mathbf{U}_f(\pi) = |0\rangle\langle 0| \otimes \mathbf{1}_S + |1\rangle\langle 1| \otimes \mathbf{S}_x \equiv \mathbf{R}_{22} \otimes \mathbf{1}_S + \mathbf{R}_{11} \otimes \mathbf{S}_x, \qquad (8)$$

which proves the assertion made above.

To find the result of the computations it is necessary to read out the state of the final density matrix $\boldsymbol{\rho}_{out} = \mathbf{U}_{comp} \boldsymbol{\rho}_{init} \mathbf{U}^{\dagger}_{comp}$. NMR methods make it possible to measure all elements of the density matrix by the state tomography [4]. As an illustration, we shall examine a variant of reading out in the particular case of a diagonal density matrix

$$\boldsymbol{\rho}_{out} = \mu_0 \mathbf{1} + \Sigma \mu_m \mathbf{P}_{mm}, \quad m=1,2,3,4 \qquad (9)$$

in the important situation where the result of the computation is one of the states $|\chi_m\rangle$, i.e., when only one of the quantities $\mu_m$ is nonzero. It is necessary to affect the real spin 3/2 with a pulsed two-frequency RF field, which generates a free-precession signal at the resonance frequencies $\Omega_{12}$ and $\Omega_{34}$ by rotating the density matrix elements by the angles $\varphi_a=\varphi_c=\pi/2$. Such a pulse corresponds to the evolution operator

$$U_3 = U_a(\pi/2)U_c(\pi/2) = (1/\sqrt{2})[\mathbf{1} + \mathbf{P}_{21} - \mathbf{P}_{12} + \mathbf{P}_{43} - \mathbf{P}_{34}] \qquad (10)$$

Under the evolution operator (10) the density matrix (9) in the Schrödinger representation takes the form

$$\rho(t) = (1/2)\{(\mu_1 + \mu_2)(\mathbf{P}_{11} + \mathbf{P}_{22}) + (\mu_4 + \mu_3)(\mathbf{P}_{33} + \mathbf{P}_{44}) +$$
$$+ (\mu_1 - \mu_2)[\mathbf{P}_{21}\exp(-it\Omega_{12}) + \mathbf{P}_{12}\exp(it\Omega_{12})] + \qquad (11)$$
$$+ (\mu_3 - \mu_4)[\mathbf{P}_{43}\exp(-it\Omega_{34}) + \mathbf{P}_{34}\exp(it\Omega_{34})]\},$$

where time is measured from the end of the computational cycle. In the state described by the density matrix (11), the quantum-mechanical averages of the transverse spin components become nonzero

$$\langle I_+(t)\rangle \equiv \langle I_x + iI_y\rangle = \langle\sqrt{3}(P_{43}+P_{21}) + 2P_{32}\rangle = \mathrm{Sp}[\rho(t)(I_x+iI_y)] =$$
$$= \sqrt{3}(\mu_3 - \mu_4)\exp(-it\Omega_{34}) + \sqrt{3}(\mu_1 - \mu_2)\exp(-it\Omega_{12}), \qquad (12)$$

The precession of the nuclear spin placed in the detection coil generates a periodic voltage at the two resonance frequencies $\Omega_{12}$ and $\Omega_{34}$ with Fourier components $\sqrt{3}(\mu_3 - \mu_4)$ and $\sqrt{3}(\mu_1 - \mu_2)$. We note that an identical pulse acting on the equilibrium density matrix (4) would produce a similar precession but with Fourier components equal to $\sqrt{3}\,Z(\lambda_3 - \lambda_4)$ and $\sqrt{3}\,Z(\lambda_1 - \lambda_2)$. A measurement of the sign of the ratios

$$b_{34} \equiv (\mu_3 - \mu_4)/(\lambda_3 - \lambda_4) \quad \text{and} \quad b_{12} \equiv (\mu_1 - \mu_2)/(\lambda_1 - \lambda_2)$$

of the corresponding Fourier components before and after the computation permits determining the final state of two virtual spins:

if $b_{34} < 0$ and $b_{12} = 0$, then the result of the computation is $|00\rangle$,

if $b_{34} > 0$ and $b_{12} = 0$, then the result of the computation is $|01\rangle$,

if $b_{34} = 0$ and $b_{12} < 0$, then the result of the computation is $|10\rangle$,

if $b_{34} = 0$ and $b_{12} > 0$, then the result of the computation is $|11\rangle$.

Nuclear spin 3/2 is not rare. Nuclei with such spin appear in the most divers and easily accessible materials. Writing of two qubits on discrete levels of a single quantum particle obviates the need for an exchange interaction between the carriers of the qubits. In existing schemes for implementing quantum gates one needs to apply the special methods in order to suppress these exchange interactions during certain time intervals. This complicate the implementation of algorithms. The scheme described above can be applied to quantum systems of arbitrary physical nature. In principle, the same purpose one can reached by using nuclear spins of large magnitude (after choosing four suitable energy levels), ESR spectra with effective spin $S^* \geq 3/2$, and optical energy levels. Only the expressions for the resonance frequencies and the matrix elements of the operators will be changed. Specifically, one can use a nuclear quadrupole resonance energy spectrum, split (or not split) by the interaction with a static magnetic field. The use of effective spins $S^* \geq 3/2$ and therefore a large number of discrete energy levels gives, generally speaking, additional possibilities, which are not discussed in the present work.